\documentstyle[aps,preprint,epsf]{revtex}
\begin{document}
\draft
\preprint{}

\title{{\it Ab Initio } Molecular Dynamics Simulation of Liquid
Ga$_x$As$_{1-x}$ Alloys}
\author{R. V.  Kulkarni and D. Stroud}
\address{
Department of Physics,
The Ohio State University, Columbus, Ohio 43210}

\date{\today}

\maketitle

\begin{abstract}

We report the results of {\it ab initio} molecular dynamics
simulations of liquid Ga$_x$As$_{1-x}$ alloys at five different
concentrations, at a temperature of $1600$ K, just above the melting
point of GaAs. The liquid is predicted to be metallic at all
concentrations between $x = 0.2$ and $x = 0.8$, with a weak
resistivity maximum near $x = 0.5$, consistent with the Faber-Ziman
expression. The electronic density of states is finite at the Fermi
energy for all concentrations; there is, however, a significant
pseudogap especially in the As-rich samples.  The Ga-rich density of
states more closely resembles that of a free-electron metal.  The
partial structure factors show only a weak indication of chemical
short-range order. There is also some residue of the covalent bonding
found in the solid, which shows up in the bond-angle distribution
functions of the liquid state. Finally, the atomic diffusion
coefficients at 1600K are calculated to be $2.1 \times 10^{-4}$
cm$^2$/sec for Ga ions in Ga$_{0.8}$As$_{0.2}$ and $1.7 \times
10^{-4}$ cm$^2$/sec for As ions in Ga$_{0.2}$As$_{0.8}$.

\end{abstract}

\newpage

\section{Introduction}

The study of liquid metals and alloys has drawn considerable attention
recently, in particular due to the possibility of carrying out first
principles or {\it ab initio} calculations for these systems. In these
calculations the electronic structure is evaluated quantum
mechanically using Density Functional Theory (DFT) and the
corresponding forces are used to move the ions according to classical
molecular dynamics. Using this approach it is possible to calculate
both the atomic and electronic structure consistently and to see how
changes in one are correlated with changes in the other. By now, a
number of groups have used these methods to calculate both the
thermodynamic and transport properties of a variety of liquid metals
and alloys\cite{kresse,holender,kirchhoff}.

Liquid semiconductors are of particular interest from the point of view
of {\em ab initio} calculations.  By ``liquid semiconductors,'' we mean liquids
of materials which are semiconducting in their {\em solid} phases, such as
Si, Ge, GaAs, and CdTe. Somewhat surprisingly, most of these are reasonably
good metals in their {\em liquid} phases.  For example, Si, Ge, and
GaAs all have conductivities near melting which lie in the
metallic range, and which tend to decrease with increasing temperatures,
as is characteristic of metals. This metallic behavior is correlated with
an increase in coordination number on melting, the liquid is thus more
close-packed than the solid and has a higher density.  
By contrast, $\ell$-CdTe is poorly conducting in its liquid state and its
conductivity {\em increases} with increasing temperature, characteristic 
of semiconductors.

Recent {\em ab initio} calculations for several of these materials
give behavior which is in good agreement with these experiments.
Godlevsky {\it et al}\cite{godlevsky} have found, in agreement with
experiment, that stoichiometric GaAs is metallic, whereas
stoichiometric CdTe is a reasonable insulator. These differences in
the electronic properties were related to the differences in the
structural properties occurring within the melt. An earlier calculation
by Zhang {\it et al}\cite{zhang} studied stoichiometric $\ell$-GaAs
using the Car-Parrinello version of {\em ab initio} molecular
dynamics\cite{cp,car}. In this calculation too, it was found that
$\ell$-GaAs is a metallic, weakly ionic liquid, with a larger
coordination number than the insulating solid phase.

In this paper, we describe a numerical study of
$\ell$-Ga$_x$As$_{1-x}$ over a range of concentrations, using {\em ab
initio} techniques. The liquid is miscible over the whole
concentration range, unlike the solid, which exists only at
stoichiometry.  Such a study is of interest for a variety of reasons.
The properties of of $\ell$-Ga and $\ell$-As stand in marked contrast
to each other. $\ell$-Ga is a close-packed liquid metal with a
coordination number of $\sim$ 9 and its electronic density of states
is almost free-electron like. In contrast, $\ell$-As has the same
coordination number as in the crystalline phase ($\sim$ 3), and is a
narrow band-gap semiconductor in the liquid state. Thus while we
expect $\ell$-Ga to show metallic bonding, $\ell$-As is expected to
retain the covalent character of the bond upon melting. Previous {\it
ab initio} calculations for these liquids \cite{li,holender} have
indeed confirmed this picture.  Thus as the stoichiometry is varied
for $\ell$-Ga$_x$As$_{1-x}$, we would expect interesting changes both
structurally and electronically. It is also of interest to compare the
structures of $\ell$-Ga$_x$As$_{1-x}$ for $x=0.2$ and $x=0.8$ with the
structure for the corresponding pure liquids (As and Ga respectively)
to see how changes in structural properties are correlated with
changes in electronic properties.

We turn now to the body of the paper.  In section II, we briefly
summarize our approach and method of calculation. Our results are
presented extensively in Section III, together with some analysis
connecting the calculations to a qualitative picture of the liquid
state.  Finally, in Section IV, we give a short concluding discussion.

\section{Method and Computational details}

Our method of carrying out the {\em ab initio} simulations is similar
to that of our previous work\cite{kulkarni1,kulkarni2}. We use the
plane-wave pseudopotential approach with generalized norm-conserving
pseudopotentials\cite{hamann} in the Kleinman-Bylander form\cite{kb},
treating the $d$-wave part as the local component. The
exchange-correlation potential is computed using the local-density
approximation (LDA), using the Ceperley-Alder result for the
exchange-correlation energy, as parametrized by Perdew and
Zunger\cite{pz}. The details of the code can be found in the
literature\cite{bockstedte,stumpf}

Our liquid-state molecular dynamics (MD) simulations were carried out
at a temperature $T = 1600$ K\cite{note1}, 
just above the melting point of GaAs ($T
= 1515$ K). We have considered five concentrations of
Ga$_x$As$_{1-x}$: x = 0.2, 0.4, 0.5, 0.6, and 0.8.  In each case, we
used a cubic 64-atom supercell with periodic boundary conditions.  For
this size cell, the actual numbers of Ga atoms in the five samples
were 13, 26, 32, 38, and 51.  (Since the minority component in the 20\% and
80\% samples, the statistics will be quite poor for them and the results
should be given greatest significance for the majority components in those
cases.)  
The atomic densities for the five
concentrations were obtained from the measured density of $\ell$-GaAs
at this temperature\cite{Glazov}, together with Vegard's Law
(i. e. linear interpolation of atomic volumes)\cite{note2}.  
We use a 10-Ry cutoff
for the energies of the plane waves included in the wave function
expansion, and $\Gamma$-point sampling for the supercell Brillouin
zone. For the electronic structure, we used Fermi-surface broadening
corresponding to an electronic subsystem temperature of $k_BT^{el} =
0.1$eV. In calculating the electronic wave functions, at each
concentration we include eight empty bands.  We control the ionic
temperature using the Nos\'{e}-Hoover thermostat \cite{nose-hoover}.
The equations of motion are integrated by means of the Verlet
algorithm, using an ionic time step of 125 a.u. ($\sim$ 3 fs).  For each
concentration, the samples were equilibrated for about 0.2ps,
following which simulations were carried out for more than 3ps.

These simulations start from an initial configuration which, for the
stoichiometric liquid, was generated from a classical molecular-dynamics
(CMD) simulation using potentials of the Stillinger-Weber form.  Although our
particular potentials were originally derived for liquid Ge$_x$Si$_{1-x}$
alloys, they should give a reasonable {\em starting} configuration for the
Ga$_{1-x}$As$_{x}$ alloys which we study.   We use these potentials but
with Ga and As masses rather than those of Si and Ge.  This model system
was melted using classical molecular dynamics at high temperatures, then
cooled down gradually to the temperature of interest.
The configuration thus obtained from CMD was equilibrated for 0.2 ps (about
double the expected relaxation time for this system).
For the nonstoichiometric liquids, we used the same starting configuration
as in the stoichiometric case, but with Ga atoms randomly substituted
for the appropriate number of As atoms (or vice versa), so as to give the
correct concentration.


\section{Results}

Fig.\ 1 shows the three partial pair correlation functions
g$_{GaGa}$(r), g$_{AsAs}$(r) and g$_{GaAs}$ for liquid
Ga$_x$As$_{1-x}$ at the concentration $x = 0.5$ and a temperature T =
1600 K.  
A number of features deserve mention. First, at
$x = 0.5$, the principal peaks of all three partial pair correlation
functions occur at about the same separation, namely 2.5\AA.  This
indicates the non-ionic character of the bonds; in ionic liquids the
partial pair correlation functions for like atoms are out of phase
with the corresponding function for unlike atoms \cite{hansen}. Our
results for g$_{\alpha \beta}$(r) are in close agreement with the recent
calculation by Godlevsky {\it et al} \cite{godlevsky}; an earlier
calculation by Zhang {\it et al} \cite{zhang} also gives the same
features although these authors get a much stronger principal peak for
g$_{GaAs}$(r) than is found in either our calculations of that of
Ref. \cite{godlevsky}. While the principal peaks are in phase, there
are some differences among the partial g(r)'s. For example, at $x =
0.5$, g$_{AsAs}$(r) has a slightly higher and narrower first peak, and
a stronger second peak, than g$_{GaGa}$(r), while g$_{GaGa}$(r) has a
broad first peak and no obvious peak beyond that.  We have also
calculated the coordination numbers for the first shell of neighbors,
defined as the integral of $4 \pi \rho r^2$g(r) from zero out to the
minimum after the first maximum in g(r).  [Here g(r) is the {\em
total} pair correlation function, which does not distinguish between
the two species, normalized so that it approaches unity at large r.]
The coordination we calculate in this way at $T = 1600$ K is 5.8, in
good agreement with the experimental estimate of 5.5 $\pm$ 0.5.
Note also that this value is larger than the
diamond-structure value of $4$ but significantly smaller than the
value expected in a close-packed liquid, which would be in the range
of 9 or 10. This value indicates the persistence of covalent bonding in
$\ell$-GaAs.

Fig.\ 2 shows g$_{GaGa}$(r) at $x = 0.8$ and g$_{AsAs}$(r) at $x =
0.2$. The latter shows more short-range order than the former -
specifically, a sharper main peak and a broad second peak, rather than
a single broad principal peak.  We believe that there are several
causes for these differences.  First, pure Ga has a much lower melting
temperature than either Ga$_{0.5}$As$_{0.5}$ or As.  Thus, at the same
temperature of 1600 K, we expect less short-range order for
g$_{GaGa}$(r) at $x =0.8$ than for g$_{AsAs}$(r) at $x = 0.2$, as seen
in our calculations.  In addition, we expect the $x = 0.2$ sample to
show some residue of the complex local structure seen in pure
$\ell$-As, which is, in turn, quite similar to that of the crystalline
phase\cite{li,bellisent}. Indeed, our calculated g$_{AsAs}(r)$ at $x =
0.2$ has some of the same features as those in the calculated g(r)
for pure $\ell$-As\cite{li}.  However, there are also some observable
differences, which may be related to the fact that, experimentally, 
pure $\ell$-As is semiconducting\cite{bellisent} 
while $\ell$-Ga$_{0.2}$As$_{0.8}$ is calculated to be
metallic (as shown later when we calculate the electronic
structure). First, if we integrate g$_{AsAs}$(r) for the $x = 0.2$
sample out to the first minimum beyond the main peak, we obtain a
coordination number of $3.2$ As atoms for the first shell of nearest
neighbors surrounding an As atom.  This is slightly larger than the
value $3$ reported for pure $\ell$-As (at T=1150 K) experimentally and
in previous calculations\cite{li,Hafner}. A more important difference
is that pure $\ell$-As has a sharp second peak in
g(r)\cite{bellisent}, whereas for g$_{AsAs}$(r) in our simulations
this second peak is very broad. While some of this broadening may
result from the higher temperature ($T = 1600 K$) in our simulations
for $x = 0.2$, these results suggest that the local structure seen in
$\ell$-As is preserved {\em only} out to the first shell of neighbors
in $\ell$-Ga$_{0.2}$As$_{0.8}$.

On the other hand, the features we see for g$_{GaGa}$(r) at $x = 0.8$
are qualitatively similar and consistent with those seen for liquid Ga
at lower temperatures \cite{holender,bellisent-funel}.  Using the
procedure indicated above, we get a Ga coordination number of 5.8 for
the first shell. If we integrate the total g(r) up to the first shell
(i.\ e., including both As and Ga neighbors of Ga), we obtain a Ga
coordination number of 6.9.  While these values are smaller than the
coordination number of 9.0 reported for pure
$\ell$-Ga\cite{bellisent-funel}, we attribute the difference to the
lowering of the first peak due to the higher temperature of our
simulations ($T=1600$ K) compared to those for pure $\ell$-Ga ($T=
702$ K and $T=982$ K) \cite{holender}.

Figs.\ 3-5 show information about the various partial alloy structure
factors at the same temperature. The partial structure factors
$S_{ij}({\bf k})$ are defined in one of the standard 
ways\cite{waseda,ashcroft}:
\begin{equation}
S_{ij}({\bf k}) = (N_{i}N_{j})^{-1/2} \langle \sum_{i} \sum_{j} 
e^{-i {\bf k}\cdot ({\bf R}_{i}- {\bf R}_{j})}\rangle - 
(N_{i}N_{j})^{-1/2}\delta_{{\bf k},0}
\end{equation}
where $i$ and $j$ denote the two components of the binary alloy.
Fig.\ 3 shows the
calculated total structure factor S(k), as weighted by
neutron scattering factors at a concentration $x = 0.5$.   
S(k) is defined by
\begin{equation}
S(k) = \frac{ {b_{i}}^2 S_{ii}(k) + 2b_{i}b_{j}S_{ij}(k) + {b_{j}}^2 
S_{jj}(k) }{ {b_{i}}^2 + {b_{j}}^2 }
\end{equation}
where $b_{i}$ and $b_{j}$ are the corresponding experimental 
neutron-scattering lengths ( $b_{Ga}$ = 7.2 and $b_{As} = 6.7$) 
For comparison, we also show in 
Fig.\ 3 the quantity S(k) as measured by neutron diffraction\cite{Bergman}.
As can be seen from the Figure, the two agree quite well. In particular, the
calculations convincingly reproduce the experimentally observed shoulder
on the high-$k$ side of the principal peak in S(k).

Fig. 4 shows the three partial alloy structure factors S$_{GaGa}$(k),
S$_{AsAs}$(k), and S$_{GaAs}$(k) at x = 0.5 and T = 1600K.  The
structure factor for like pairs is always positive, with a conspicuous
first peak, while the structure factor between opposite pairs is
negative for small k, becoming positive at k values corresponding to
the peaks in the other two partial structure factors.  It is of
interest to compare these results to those found in other model
calculations.  For a mixture of hard spheres of packing fraction 0.45
(characteristic of the liquid near melting) and ratio of hard sphere
diameters of 0.9\cite{ashlang}, the cross-correlation function is
negative at small k and has a peak near that of the two same-species
functions, as in our calculations.  By contrast, for liquid NaCl,
which is strongly ionic), the cross-correlation function has a strong
{\em negative} peak at the same k as the peaks of the same-species
partial structure factors.  Thus, our results are more similar to the
hard-sphere structure factors, suggesting that liquid GaAs is at most
only weakly ionic.

Finally, Fig.\ 5 shows S$_{GaGa}$(k) at $x = 0.8$ and
S$_{AsAs}$(k) at $x = 0.2$.  Once again, like the real-space
correlation functions at the same concentrations, the Ga-Ga structure
factors show slightly less correlation (i.\ e., a lower principal peak
and a less conspicuous second peak) than do the corresponding As-As
structure factors.  One possible reason for this behavior, as for the
corresponding real-space correlation functions, is that the $x = 0.8$
sample is further from melting than is the $x = 0.2$ liquid.

S$_{AsAs}$(k) for $x=0.2$ shows characteristic differences from
that for pure $\ell$-As, which are analogous to those discussed
earlier for the g(r)'s.  Specifically, S(k) for pure
$\ell$-As\cite{bellisent} has a split principal peak with maxima at $
k= 2.45$ $\AA^{-1}$ and $k=3.74$ $\AA^{-1}$.  By contrast,
S$_{AsAs}$(k) for $x = 0.2$ has a peak at 2.5 $\AA$, but the
second peak is reduced to only a shoulder at about 3.5 $\AA$.
The fact that the second of the split peaks is smoothed to a
shoulder at $x=0.2$ indicates a change in the local structure which is also
reflected in the reduction of the second peak of partial g(r) 
as noted earlier.

We next discuss the $x=0.8$ sample, comparing our results with those
for pure $\ell$-Ga at lower temperatures
\cite{holender,bellisent-funel}. The partial structure factor show the
same qualitative features as the structure factors for the pure
liquid, but there are some quantitative differences. Our results for
$x = 0.8$ (at $1600$ K) show a first peak in S$_{GaGa}$(k) with
a maximum of only about 1.3,which is lower than the experimental
one seen in pure Ga at $959$ K ($\sim$ 1.7). We attribute the lowering
of the first peak to the increased temperature, as seen experimentally
in most liquid metals.

Further information about the short-range order in the liquid alloy
may be obtained from {\em bond angle distribution functions}, shown in
Figs.\ 6 and 7.  These functions are defined by analogy with our
previous work in liquid Ge\cite{kulkarni1,kulkarni2}.  Namely, one
considers a group of three atoms.  Of these, one is denoted as the
central atom; the other two atoms (denoted as ``side atoms''), with
the central atom, define a bond angle $\theta$.  $g_3(\theta,r_c)$ is
the distribution of bond angles formed by all such groups of three
atoms, such that both the side atoms lie within a cutoff distance
$r_c$ of the central atom.  Fig.\ 6 shows $g_3(\theta, r_c)$ for the
Ga-Ga-Ga angles at $x = 0.8$, and for the As-As-As angles at 
$x = 0.2$, each for two different choices of the cutoff radius $r_c$.
Fig.\ 7 shows the same functions for Ga-Ga-Ga and for As-As-As at 
$x =0.5$.

The peaks in these distribution functions give hints about about the
short-range bond-order in the liquids.  For example, a peak near
$\theta = 60^o$ corresponds to a relatively close-packed arrangement
of the corresponding atomic group, with many nearest neighbors.  In
contrast, a peak near $100^o$ indicates a more tetrahedral structure,
typical of covalent bonding. Thus, the upper part of Figs.\ 6 and 7
suggest that the Ga ions form a rather close-packed arrangement at $x
= 0.8$ and $x = 0.5$, since there is a strong peak near $60^o$ for
both values of the cutoff $r_c$. By contrast, the lower parts of
Figs.\ 6 and 7 suggest that, as expected, the As atoms have a more
open arrangement at $x = 0.2$, since there are {\em two} peaks in
$g_3$ at this concentration: a side peak near $50^o$, and another
noticeable peak near $97^o$. We also observe the peak at $97^o$ for As
atoms at $x=0.5$, however it is broader and less pronounced than at
$x=0.2$.  In both Figs.\ 6 and 7, the Ga-Ga-Ga bond angle distribution
depends little on $r_c$.

At $x = 0.2$, the As-As-As distributions show noticeable $50-60^o$
peaks only at the larger cutoff values.  At the smaller cutoff radius,
the $50-60^o$ peak is missing. Thus, at short distances, the As clusters
tend to maintain the local version of the structure they have in the
pure liquid phase (and the crystalline phase) which shows a strong
peak at $97^o$, but at larger cutoffs the local structure differs from
that of pure $\ell$-As.  We have made the same observation in
connection with g$_{AsAs}$(r) at the same concentrations, and with the
shoulder in S$_{AsAs}$(k). The $97^o$ peak implies some tetrahedral
order persisting to $x = 0.2$, though this peak is less pronounced
than in pure $\ell$-As\cite{li}.

We have also calculated the electronic properties of 
$\ell$-Ga$_x$As$_{1-x}$.  We calculate the single-particle electronic
density of states $N(E)$ in the standard way, by using the expression
\begin{equation}
N(E) = \sum_{{\bf k},E_{{\bf k}}}w_{{\bf k}}g(E-E_{{\bf k}}).
\end{equation}
In this expression $E_{{\bf k}}$ denotes one of the energy eigenvalues
of the single-particle wave functions at a particular ${\bf k}$ point
within the supercell Brillouin zone, w$_{{\bf k}}$ is the weight of
that ${\bf k}$ point (defined below), and g(E) is a Gaussian smoothing
function of width $\sigma = 0.2eV$. Our calculation is carried out by
sampling the supercell Brillouin zone at eight special ${\bf k}$
points, using the same choice of special points and weights as that of
Holender {\it et al} in their calculations for pure
$\ell$-Ga\cite{holender}, which have been well tested and found to be
an adequate representation of the supercell Brillouin zone.  This choice
is, however, convenient rather than unique; we expect that other choice having
the same number of {\bf k} points would have given similar results, as has
been found in pure $\ell$-Ga.  
For each ${\bf k}$-point we include 40
conduction band states, and for each concentration, we obtain our
final results by averaging over twelve representative liquid state
configurations.

The resulting calculated density of states N(E) is shown for the four
concentrations $x = 0.2, 0.4, 0.6$, and $0.8$ in Fig.\ 8.  [We do not
show our calculated N(E) for $x=0.5$, but it interpolates smoothly
between $x=0.4$ and $x = 0.6$.]  As in our previous studies, the alloy
has a clearly metallic density of states for all concentrations $x$.
However, just as in our previous results for Ga$_x$Ge$_{1-x}$, the
density of states becomes more and more free-electron like as the
concentration $x$ of the metallic component (Ga in this case)
increases.  Pure $\ell$-As is semiconducting and has been calculated
to have deep minima in the electronic density of states at the Fermi
energy ($E_F$), and also at an energy of $\sim$ -7 eV (measured from
$E_F$) \cite{li}, which separates the $s$ and $p$ bands. Liquid Ga, on
the other hand, has an almost free-electron like density of states.
We see these features reflected in our simulations; as the figures
indicate, the electronic density of states (DOS) has a pseudogap in
the As rich phase which progressively fills up as the Ga concentration
is increased, so that for $x=0.8$ it is hardly noticeable.  But even
at low Ga concentration ($x=0.2$), there is no minimum in the density
of states at the Fermi energy. As for the pseudogap, we find that its
position changes monotonically to lower energies (relative to $E_F$)
with increasing Ga concentration. A similar pseudogap is reported in
calculations for pure $\ell$-Ge \cite{kulkarni1,jank}, for the same
reason, i. e., a partial separation between $s$-like and $p$-like
bands.

We have also computed the frequency-dependent electrical conductivity
$\sigma(\omega)$ for $x = 0.2$, $0.4$, $0.5$, $0.6$, and $0.8$, at
frequencies ranging up to 2eV.  $\sigma(\omega)$ is given by the
standard Kubo-Greenwood expression\cite{thouless}
\begin{equation}
\sigma(\omega ) = \frac{2 \pi e^{2}}{3m^{2} \omega \Omega } \sum_{i} \sum_{j} 
\sum_{\alpha}
 (f_{j} - f_{i}) |\langle \psi_{i}| \hat {p_{\alpha }} |\psi_{j} \rangle |^{2}
 \delta(E_{j} -
 E_{i} - \hbar \omega ).
\end{equation}
where $\psi_{i}$ and $\psi_{j}$ are the single particle Kohn-Sham wave
functions with Fermi occupancies $f_{i}$ and $f_{j}$ and energy
eigenvalues $E_{i}$ and $E_{j}$. Once again, we calculate the
conductivity using the same set of eight special {\bf k} points used
for N(E), and averaging over the same twelve representative ionic
configurations, including 40 conducting band states for each ${\bf
k}$.  While there is some inaccuracy in using differences in 
Kohn-Sham eigenvalues, as we do in this equation, it is difficult to do 
better at present in liquid metals and semiconductors, and this same 
approach does give good agreement with experiment in a number of other
liquid metals, especially at zero frequency but also for stoichiometric
$\ell$-GaAs\cite{godlevsky} at finite frequencies.

The frequency-dependent conductivity is shown in Fig.\ 9 for $x =
0.2$ and $x = 0.8$, and its calculated zero-frequency limit is listed
in Table I for the five concentrations x = 0.2, 0.4, 0.5, 0.6 and 0.8,
all at $T = 1600K$.  

Several features of the conductivity graphs, and of the tabulated d.\
c.\ limits, deserve mention.  First, the calculated value of the d.\
c.\ conductivity at x = 0.5 is very close to the experimental value:
$0.84 \times 10^4$ ohm$^{-1}$cm$^{-1}$, compared to the experimental
value of $0.79 \times 10^4$ ohm$^{-1}$cm$^{-1}$\cite{Glazov}.
Secondly, the conductivity has a weak minimum near x = 0.5.  This is
consistent with expectations based on second-order perturbation
theory\cite{faberziman} which would predict that alloy scattering (due
to concentration fluctuations) would be a {\em maximum} near x = 0.5
Thirdly, the frequency-dependence of $\sigma(\omega)$ becomes more
metallic as $x$ increases.  Specifically, at x = 0.8, $\sigma(\omega)$
clearly decreases with increasing frequency, characteristic of a Drude
metal, while for the highest As concentration (x = 0.2), the
conductivity is nearly frequency-independent in the range of
calculation.  This behavior is closer to the nonmetallic behavior in
which the conductivity {\em increases} with increasing frequency.

Finally, we have computed one important {\em atomic} transport
coefficient, namely, the atomic self-diffusion coefficients $D_{ii}$
for the majority species in the two liquids Ga$_{0.2}$As$_{0.8}$ and
Ga$_{0.8}$As$_{0.2}$.  In both cases, the $D_{ii}$'s can be extracted
from a plot of the mean-square atomic displacement versus time, which
approaches a straight line in the limit of large time.  The expression
is
\begin{equation}
D_{ii} = \lim_{t \rightarrow \infty}
\frac{\langle |{\bf R}_{i}(t)-{\bf R}_{i}(0)|^2 \rangle}{6t},
\end{equation}
where ${\bf R}_{i}(t)$ denotes the position of an ion of species 
$i$ at time $t$, and the triangular brackets denote an average over
all atoms of species $i$ and over initial times.  Plots of the
mean-square atomic displacement as a function of time are shown in
Fig.\ 10 for both types of atoms at $x = 0.2$ and $0.8$ as indicated above;
>From these we obtain $D_{Ga} = 2.1 \times 10^{-4}$ cm$^2$/sec at 
$x = 0.8$; $D_{As} = 1.7 \times 10^{-4}$ cm$^2$/sec at $x = 0.2$. 

The value for $D_{As}$ at $x = 0.2$ is about three times larger than
that obtained by Li \cite{li} for pure $\ell$-As ($D \sim 0.6 \times
10^{-4}$ cm$^2$/sec).  This difference is probably due to a
combination of several factors.  First, the calculations by Li are
carried out at a temperature $\sim$ 450 K lower than ours.  Secondly,
pure $\ell$-As seems to have more covalent bonding than
Ga$_{0.2}$As$_{0.8}$, which probably impedes atomic motion, giving a
lower atomic diffusion coefficient for As. Thus, in short, this
behavior seems to be consistent with the rest of our picture, which is
that $\ell$-Ga$_x$As$_{1-x}$ rapidly acquires metallic conductivity,
and corresponding structural properties for $x$ as small as 0.2.

\section{Discussion and Conclusions}

The most striking results of these calculations is that
Ga$_x$As$_{1-x}$ remains metallic at all concentrations between $x
=0.2$ and $x = 0.8$, including the As rich value $x = 0.2$. Thus even
low Ga concentrations are sufficient to render $\ell$-Ga$_x$As$_{1-x}$
metallic (recall that pure $\ell$-As is semiconducting). This is
reflected in the electronic density of states which shows no minimum
at the Fermi energy for all concentrations studied.  The liquid
structure is also consistent with metallic behavior at all
concentrations between $x=0.2$ and $0.8$, although there are some
clear deviations from the behavior seen in simple metallic alloys.
Specifically, although the coordination number at all concentrations
is larger than the value of four that might be expected in a
predominantly covalent liquid, it is still smaller than that of a
typical close-packed hard-sphere mixture.  At $x = 0.2$ and $0.8$, the
pair correlation functions and structure factors resemble those of the
corresponding pure liquids, except that the split first peak in $S(k)$
of $\ell$-As becomes a single peak with a weak shoulder in As.

It has been suggested \cite{joffe} that semiconducting properties
persist in a liquid only if the liquid has the same short-range order
as the crystalline phase.  In the alloys we study, the liquid state
has a short range order which is distinctly different from the solid.
We also find that all these alloys are metallic, in agreement with the
suggestion of Ref.\ \cite{joffe}.  By contrast, again in agreement
with the picture advanced in \cite{joffe}, another compound
semiconductor, stoichiometric CdTe, appears to preserve the
crystalline short-range order, and also to retain its semiconducting
characteristics in the liquid state. Such semiconducting behavior was
indeed found in the {\em ab initio} calculation by Godlevsky {\it et
al}\cite{godlevsky}.  It would be of interest to extend their
calculations off stoichiometry, where metallic behavior is likely.

The other characteristics of $\ell$-Ga$_x$As$_{1-x}$ reflect its
fundamentally liquid-metallic character.  For example, the resistivity
is predicted to exhibit typical Faber-Ziman behavior: a weak positive
deviation from a linear concentration dependence near stoichiometry,
which is caused by alloy scattering.  The calculated values of the
partial atomic diffusion coefficients are comparable to the diffusion
coefficients of those other liquid semiconductors which are metallic
in their liquid states.  We find no evidence of a strong reduction in
this value because of formation of clusters near stoichiometry; such
cluster formation is not expected for Ga$_x$As$_{1-x}$ because of the
small electronegativity differences between the two species.  One
might speculate that, in other liquid semiconductors, such as
stoichiometric $\ell$-CdTe, which remain poorly conducting in the
liquid state, the local structure is much more ionic near melting and
the atomic diffusion coefficients are correspondingly lower.

In summary, our calculations show that Ga$_x$As$_{1-x}$ is a
reasonable metal at all concentrations between $x = 0.2$ and $x =
0.8$.  In particular, there is no evidence of strong compound
formation in the liquid state near $x = 0.5$. The electrical
conductivity shows a concentration dependence typical of a liquid
metallic alloy, with evidence of weak scattering from concentration
fluctuations which reaches a maximum near $x = 0.5$. The electronic
density of states shows no minimum at the Fermi energy; instead, it
has a pseudogap between $s$-like and $p$-like occupied state which
persists at all concentrations, though it is considerably weaker in
the Ga-rich alloys.  The atomic diffusion coefficient is calculated to
be similar to that of other liquid semiconductors which are metallic
in their liquid state.  Finally, the liquid structure shows some
indications of deviation from the behavior of simple liquid metal
alloys\cite{ashcroft}.  The principal evidence of deviation from
behavior characteristic of a simple liquid alloy formation comes from
the calculated pair correlation functions, structure factors, and bond
angle distribution functions, all of which show some weak indications
of departure from close-packed behavior: smaller coordination
functions than in typical hard-sphere liquids, and a weak residue of
tetrahedral bonding.

\section{Acknowledgements}

We are grateful for support by NASA, Division of Microgravity Sciences,
under grants number NCC3-555 and NCC8-152.
We also thank M. Scheffler and the research group at the
Fritz-Haber Institute (Theory Department) for use of their
{\em ab initio} code FHI96MD, and D.\ Matthiesen, A.\ Chait, and
E.\ Almaas for many valuable conversations.   
These calculations were carried out
using the SP2 at the Ohio Supercomputer Center and the RS/6000 cluster at the
NASA Lewis Research Center.

\newpage

\newpage

\begin{table}
\caption{D.\ c.\ conductivity at the five concentrations,
obtained by extrapolating
low frequency a.\ c.\ conductivity results.}
\begin{tabular}{|l|l|l|l|l|l|} \hline
Concentration ($\%$)        &   20   &   40      & 50 & 60   &   80  \\ \hline
${\sigma_{dc}}$ (10$^{4}$ ohm$^{-1}$ cm$^{-1}$) &0.93   & 0.9 & 0.84 & 0.91   
& 1.1   \\  \hline
\end{tabular}
\end{table}

\vspace{0.2in}

\newpage

\begin{center}
{\bf FIGURE CAPTIONS}
\end{center}


\begin{enumerate}

\item (a) Pair correlation functions g$_{GaGa}$(r), g$_{AsAs}$(r),
and g$_{GaAs}$(r) 
for $\ell$-Ga$_x$As$_{1-x}$ at a temperature of 1600 K and
concentration $x = 0.5$. The graphs are vertically offset by one unit each
for clarity.

\item (top) g$_{GaGa}$(r) for x = 0.8; (bottom) g$_{AsAs}$(r) at x = 0.2
and $T=1600$ K.

\item Full line: Calculated neutron structure factor  
S(k) = $\sum_{i,j = Ga}^{As}$f$_{ij}$(k)S$_{ij}$(k),
where $f_{ij}(k)$ is the  neutron form factor, 
for $x = 0.5$.
Open circles: measured S(k) as obtained by neutron diffraction 
[ref.\ \cite{Bergman}].

\item Partial structure factors S$_{GaGa}$(k), S$_{AsAs}$(k), and
S$_{GaAs}$(k) for $x = 0.5$. S$_{AsAs}$(k) is vertically offset for 
clarity.

\item Calculated partial structure factors (top) S$_{GaGa}$(k) at x = 0.8;
(bottom) S$_{AsAs}$(k) at x = 0.2.

\item Calculated bond angle distribution functions $g(\theta, r_c)$ for (top)
groups of three Ga atoms at a concentration $x = 0.8$, and (bottom) groups
of three As atoms at a concentration $x = 0.2$, both 
for two different cutoff radii, $r_c = 3.4 \AA$
and $r_c = 3.8 \AA$, as defined in the text.  

\item Same as for Fig.\ 6 but for x = 0.5 and
(top) g$_{\mbox{GaGaGa}}(\theta, r_c)$; 
(bottom) g$_{\mbox{AsAsAs}}(\theta, r_c)$;

\item Single-particle electronic density of states $N(E)$ for Ga$_x$As$_{1-x}$
at $x = 0.2$, $0.4$, $0.6$, and $0.8$.
The Fermi energy in each case is shown as a dashed vertical line.

\item Calculated electrical conductivity $\sigma(\omega)$ for 
Ga$_x$As$_{1-x}$ at $x = 0.2$ and $x = 0.8$.

\item Calculated mean-square displacement 
$\langle|{\bf R}_{i}(0) - {\bf R}_{i}(t)|^2\rangle$, plotted
as a function of $t$ at x = 0.2 for As and at x = 0.8 for Ga.  
  
\end{enumerate}

\begin{figure}[tb]
\epsfysize=16cm
\centerline{\epsffile{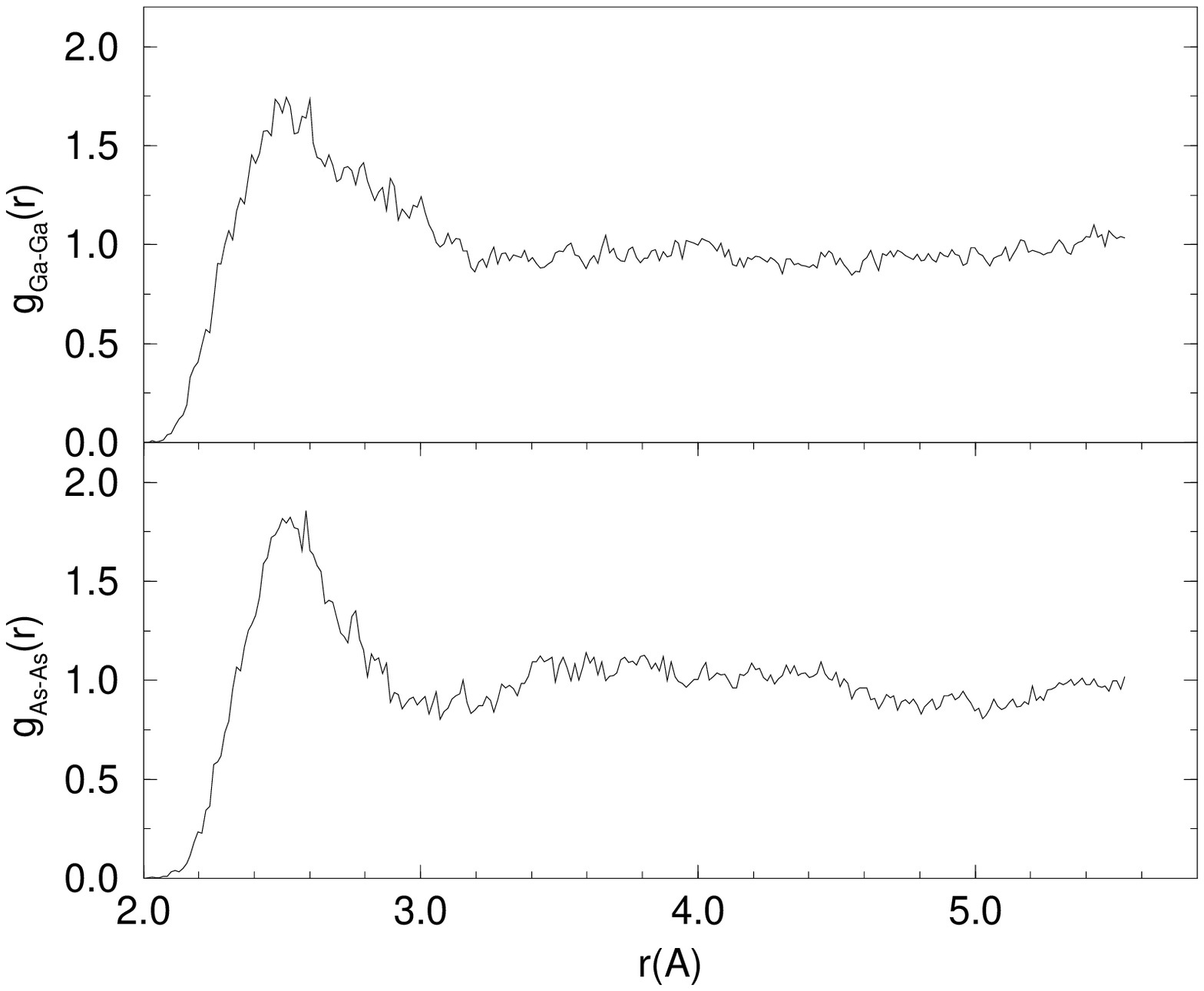}}
\caption{}
\end{figure}
\newpage

\vspace*{3cm}
\begin{figure}[tb]
\epsfysize=16cm
\centerline{\epsffile{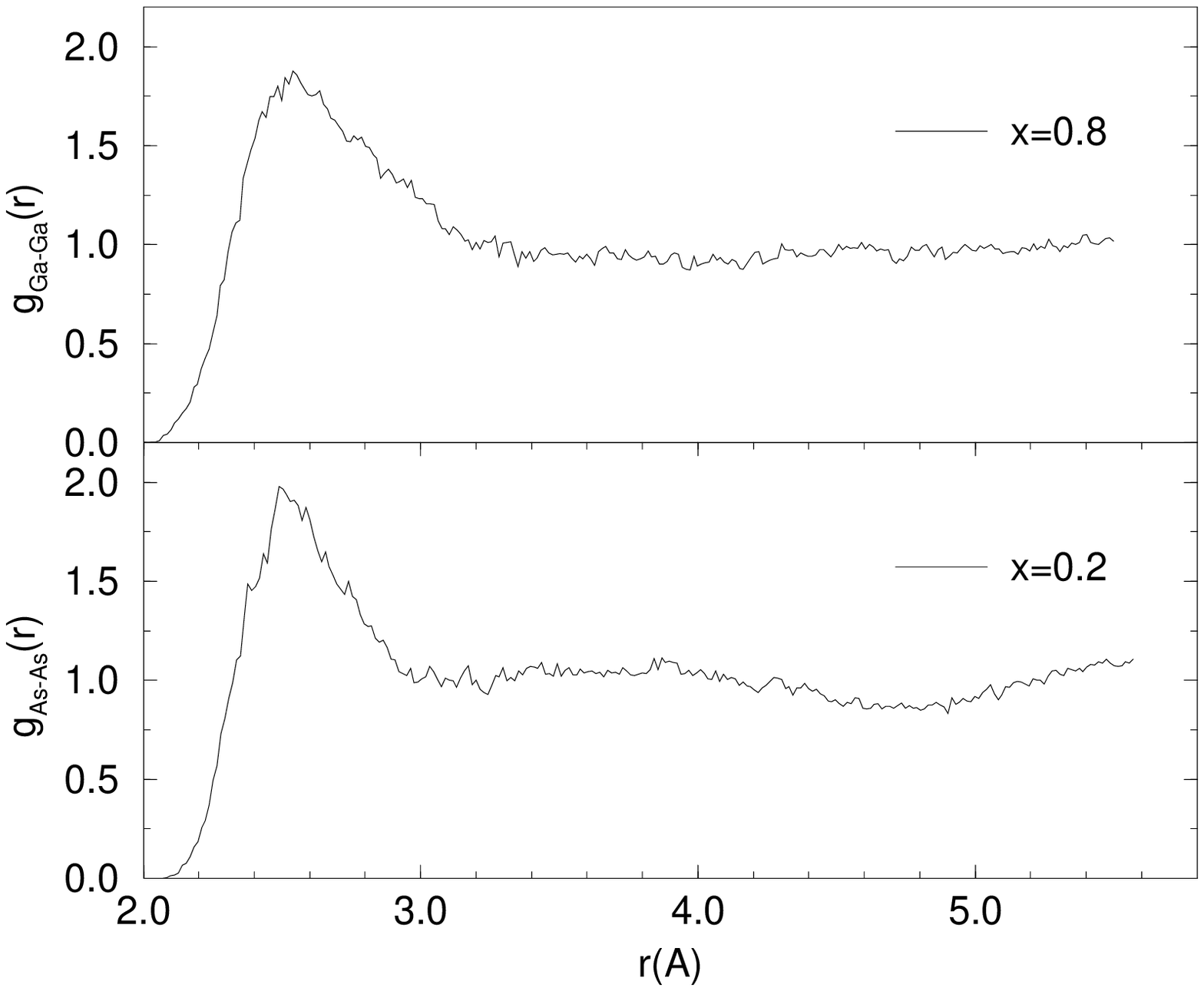}}
\caption{}
\end{figure}

\newpage

\vspace*{3cm}
\begin{figure}[tb]
\epsfysize=16cm
\centerline{\epsffile{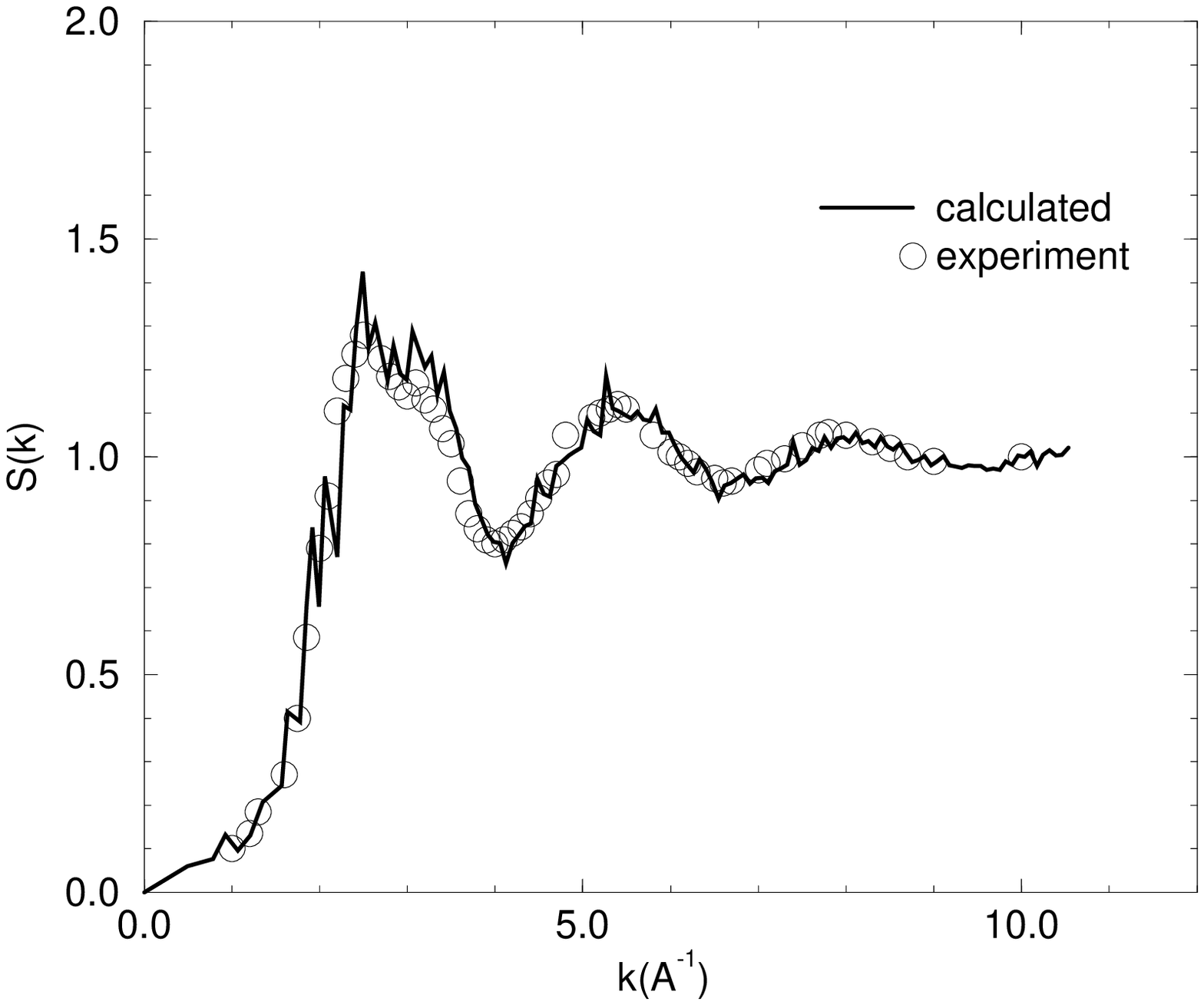}}
\caption{}
\end{figure}

\newpage

\vspace*{2.5cm}
\begin{figure}[tb]
\epsfysize=16cm
\centerline{\epsffile{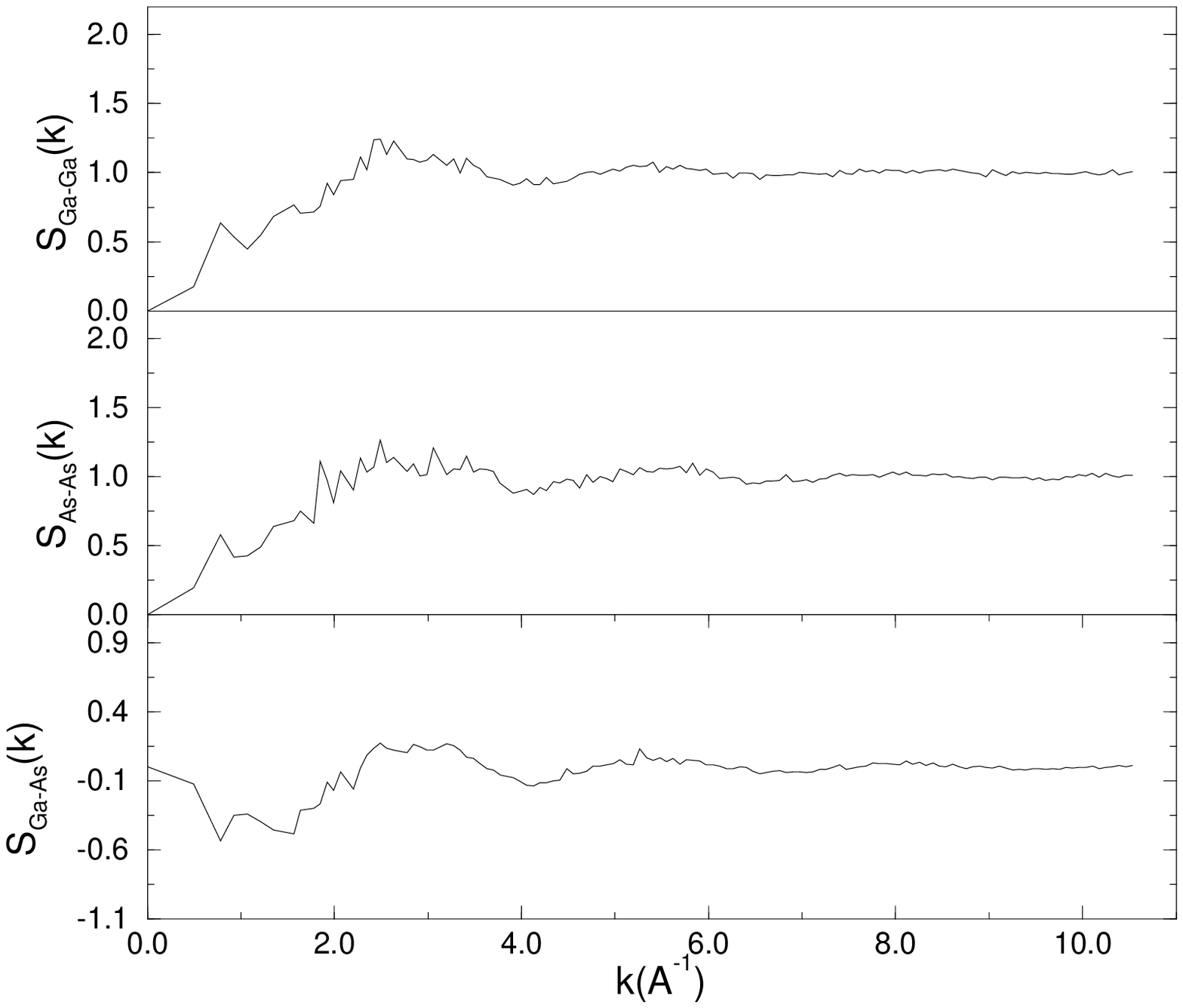}}
\caption{}
\end{figure}

\newpage

\vspace*{3cm}
\begin{figure}[tb]
\epsfysize=16cm
\centerline{\epsffile{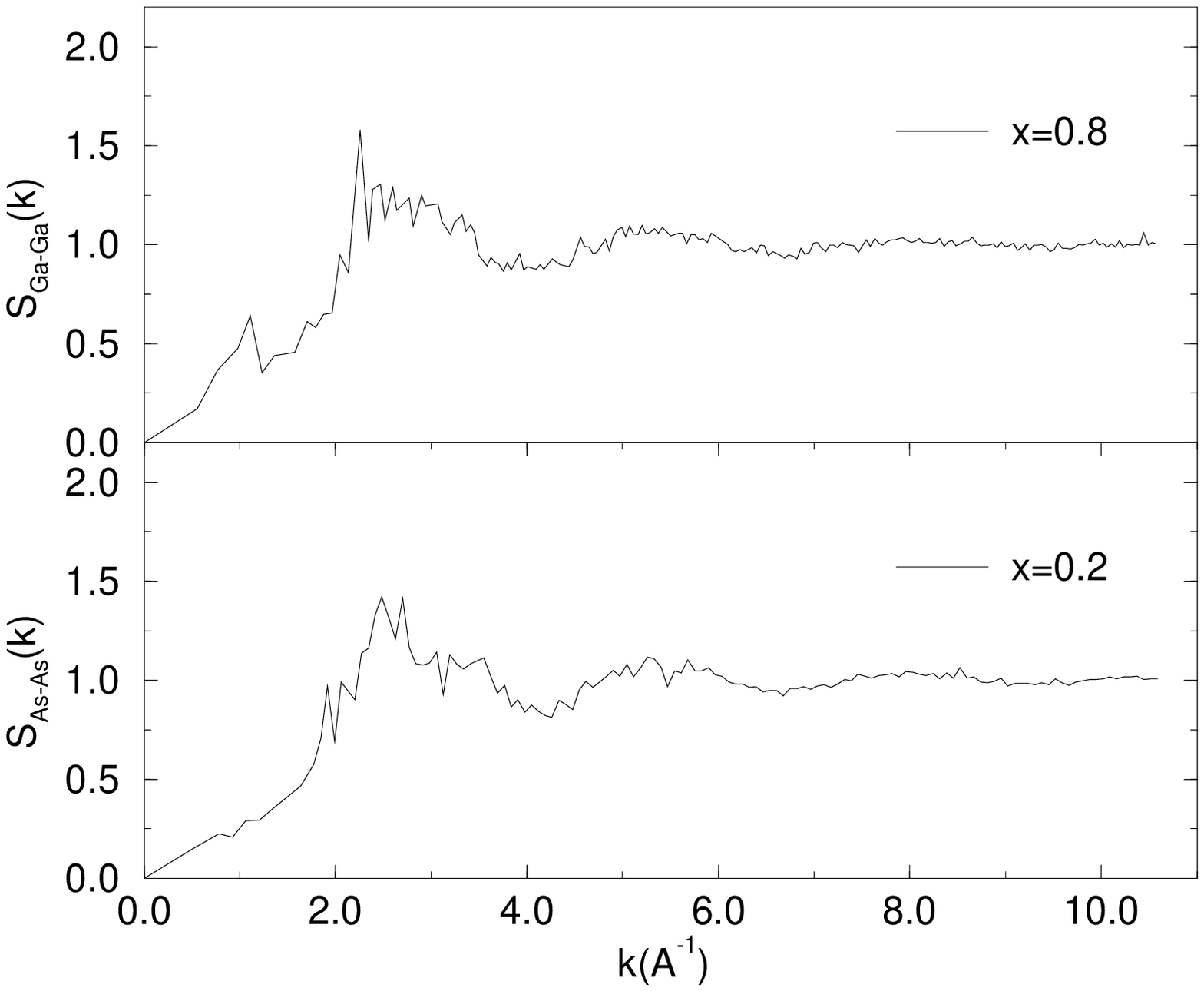}}
\caption{}
\end{figure}

\newpage

\vspace*{3cm}
\begin{figure}[tb]
\epsfysize=16cm
\centerline{\epsffile{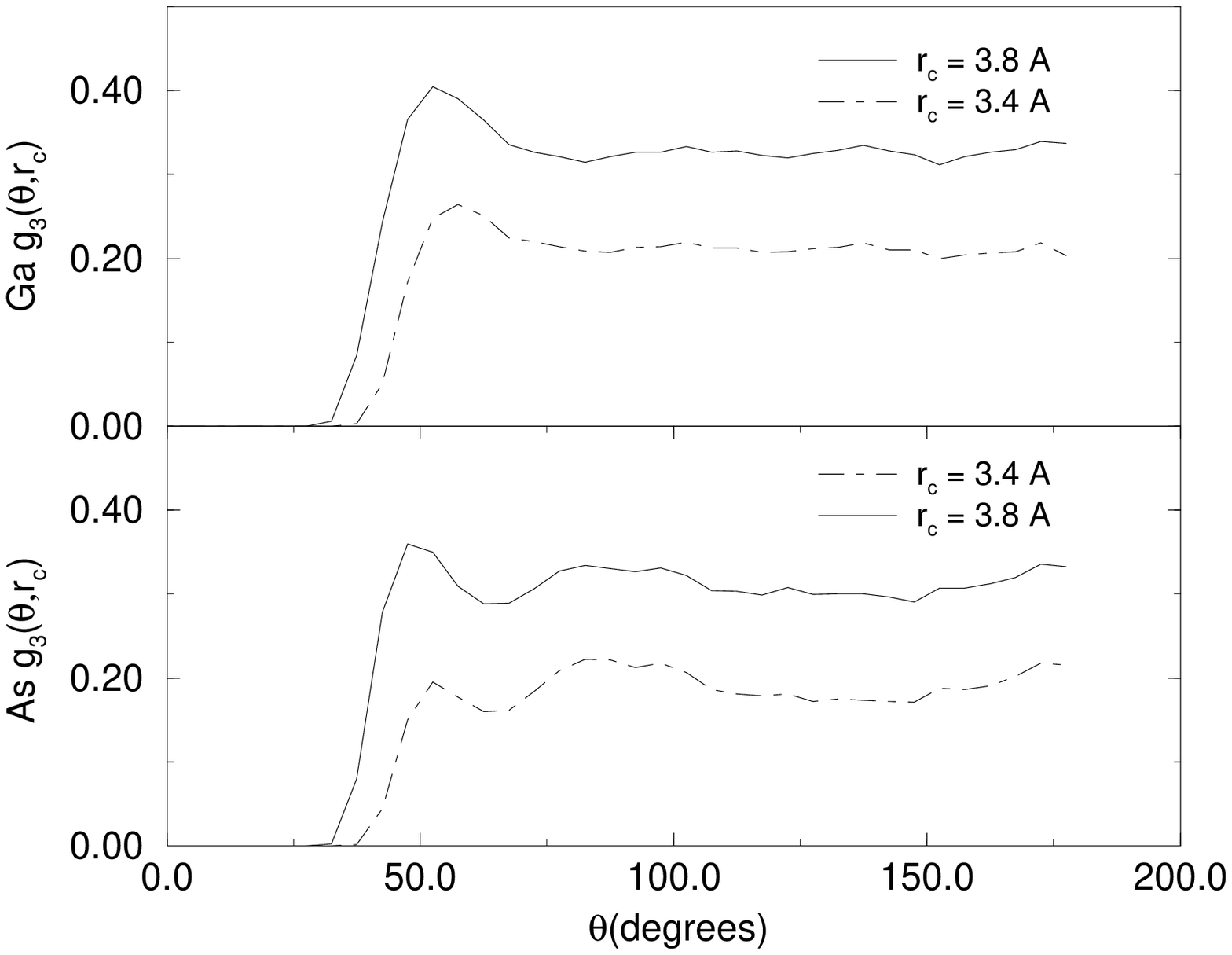}}
\caption{}
\end{figure}

\newpage

\vspace*{3cm}
\begin{figure}[tb]
\epsfysize=14cm
\centerline{\epsffile{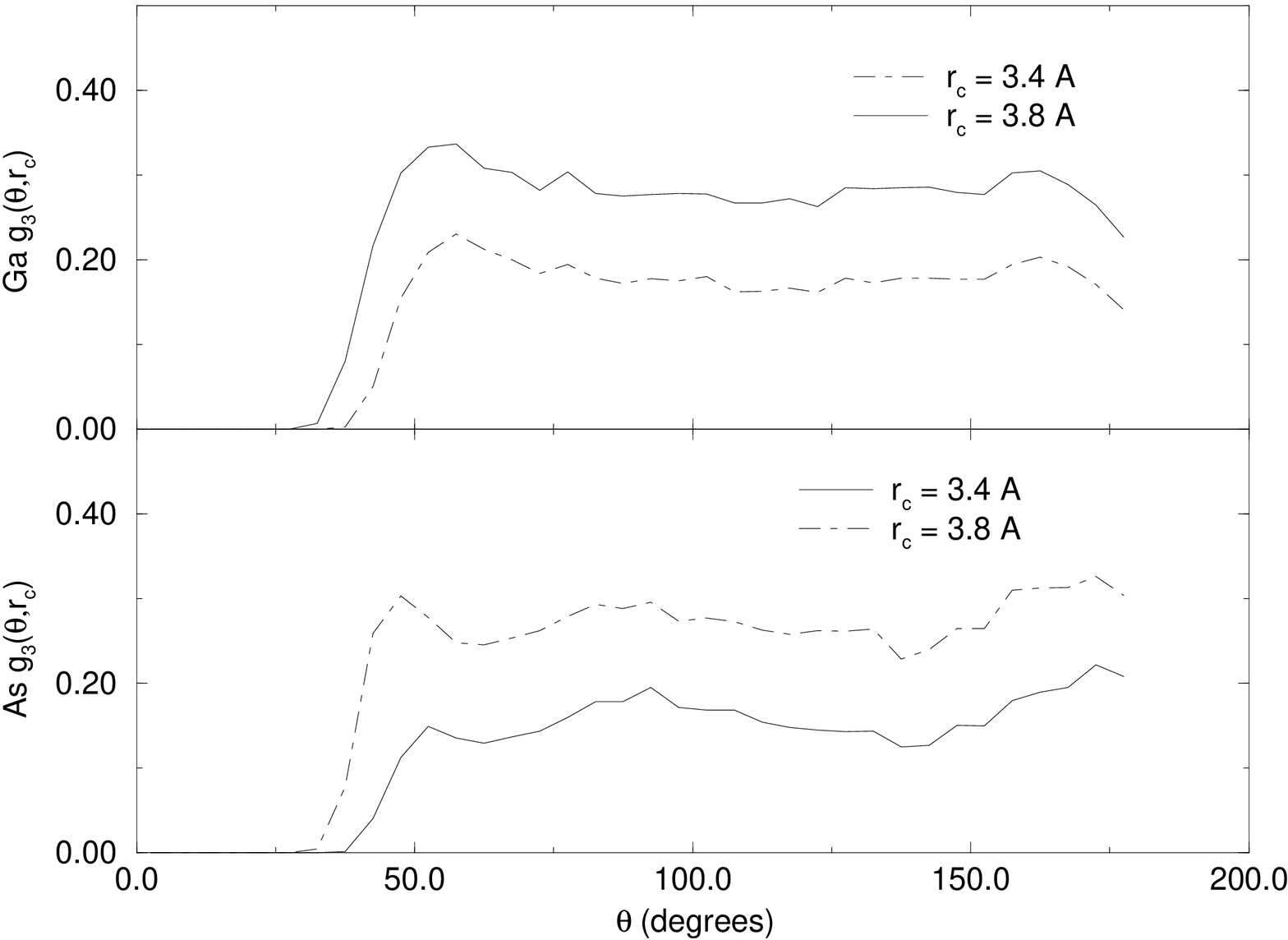}}
\caption{}
\end{figure}

\newpage

\vspace*{3cm}
\begin{figure}[tb]
\epsfysize=15cm
\centerline{\epsffile{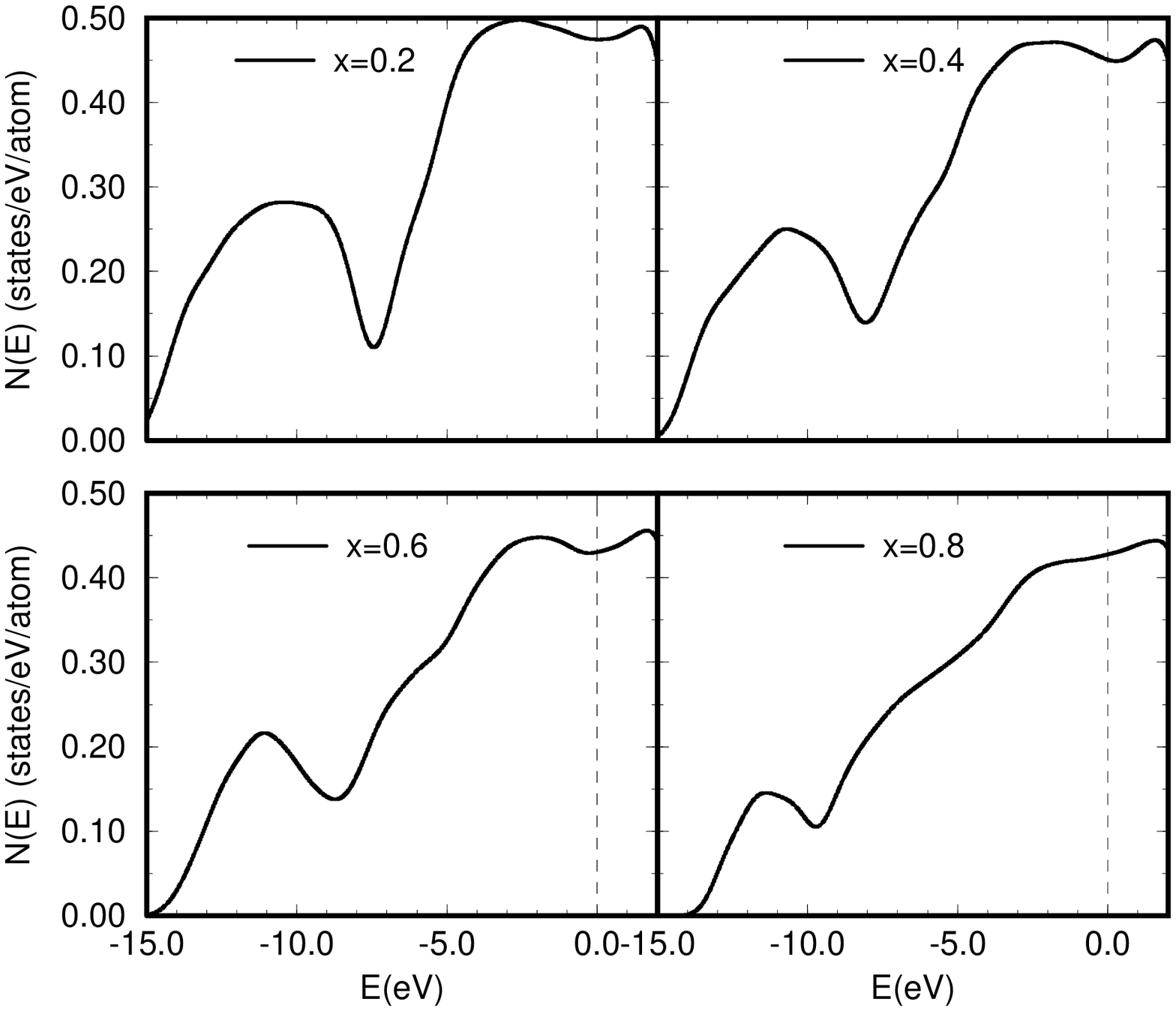}}
\caption{}
\end{figure}

\newpage

\vspace*{3cm}
\begin{figure}[tb]
\epsfysize=16cm
\centerline{\epsffile{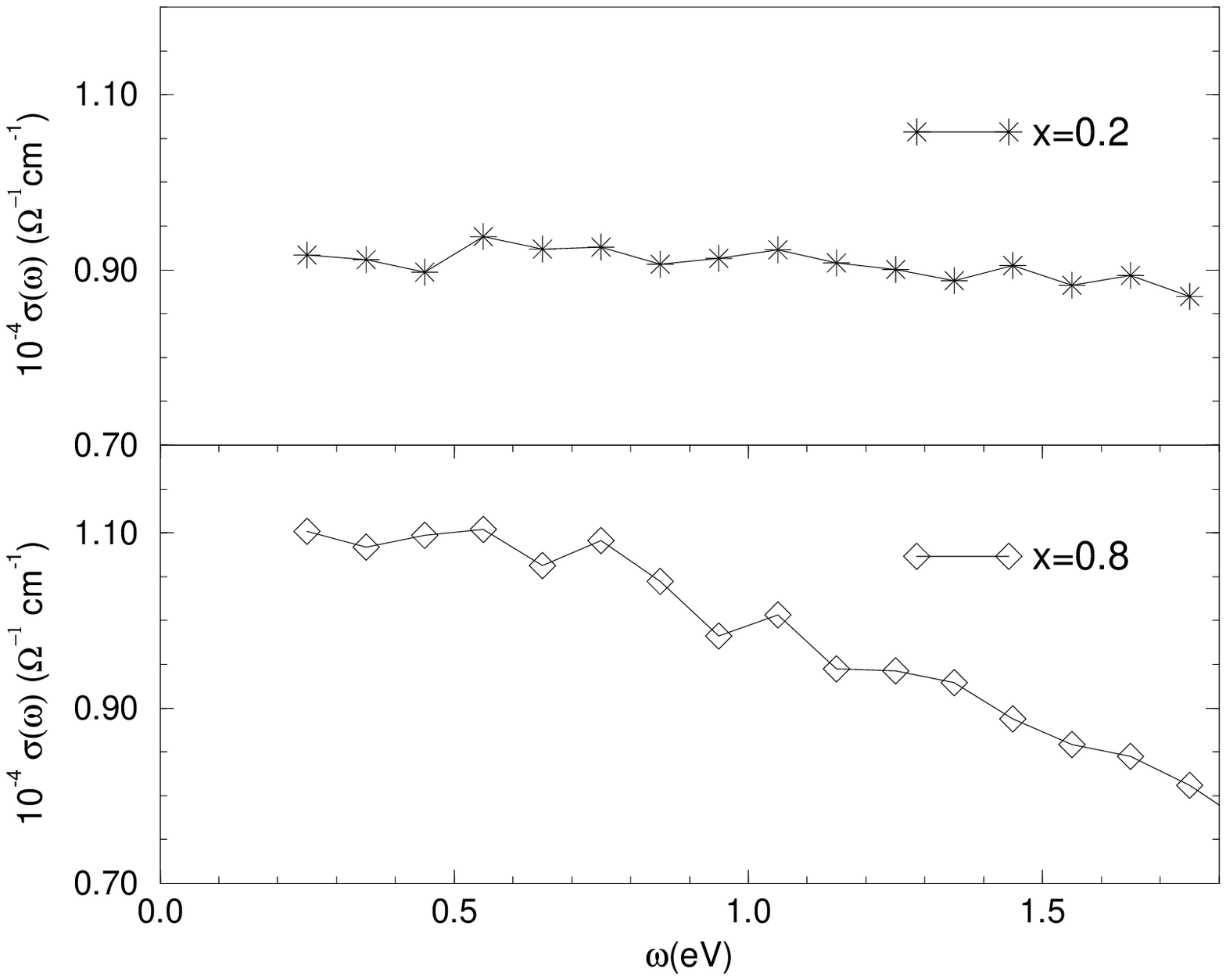}}
\caption{}
\end{figure}

\newpage

\vspace*{3cm}
\begin{figure}[tb]
\epsfysize=16cm
\centerline{\epsffile{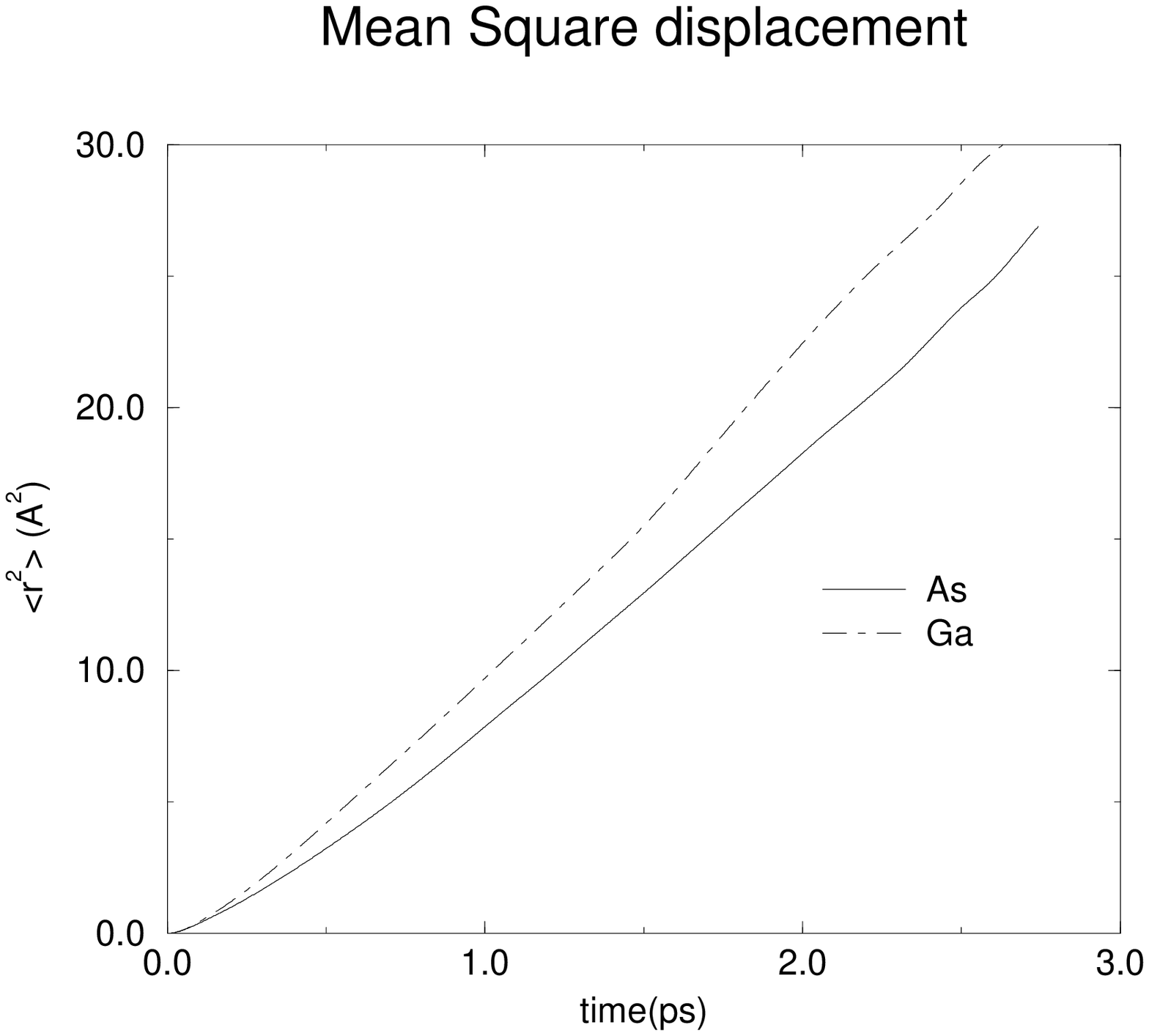}}
\caption{}
\end{figure}


\begin{references}
\bibitem{note1} Of course 1600K is much closer to the melting point of
stoichiometric GaAs and of pure As than it is to that of metallic Ga. 
Nontheless, a calculation at constant temperature seems the best way to
make uniform comparisons.
\bibitem{note2} Vegard's Law may well be inaccurate for
$\ell$-Ga$_x$As$_{1-x}$, but in the absence of experimental data
for the concentration-dependence of the volume, it seems a reasonable
interpolation.
\bibitem{note1} For the partial correlation functions $g_{ij}(r)$
and $S_{ij}(k)$, we use the standard definitions as given, for
example, in Ref.\ (\cite{waseda}) or Ref.\ (\cite{ashcroft}).
\end{references}
\end{document}